# Laboratory-Scale Simulation of Spiral Plumes in the Mantle

*A.N. Sharifulin, A.N.Poludnitsin**

*Perm National Research Polytechnic University, 614600 Perm, Russia*
**Perm State National Research University, 614600 Perm, Russia*

On the basis of laboratory simulation a mechanism is established for the formation of the upper mantle convection spiral plumes from a hot point in the presence of a roll-type large-scale convective flow. The observed plume has horizontal sections near the upper limit, which may lead to the formation of chains of volcanic islands.

**Introduction.** The hypothesis of a vertical form of convective plumes in the mantle is implicitly introduced in [1]. The most vividly it is articulated in [2], where mantle plumes are referred to as "columns of heat". The hypothesis that mantle plumes can be inclined due to a shear-flow beneath tectonic plates has been developed in [3]. Additionally, it was shown that convective instability of horizontal regions of such plumes may be a reason for secondary plumes, leading to formation of chains of volcanic islands such as Hawaii archipelago. The laboratory simulation of the horizontal part of the plume in this paper by artificially forming an inclined jet of less viscous fluid in the reservoir with very viscous silicon oil has been done. The study of the interaction of the thermal plume and the large-scale shear flow for the understanding of processes in the mantle has not been carried out.

In this paper we experimentally simulated the appearance of a plume from the hot spot and study its interaction with cellular flow, simulating beneath the plates shear flow. It is shown that the presence of cellular convective motion may lead to the formation of a spiral convective plume.

**Experimental setup.** The mantle substance is characterized by extremely high value of the Prandtl number $Pr \approx 10^{24}$ [4,5], so the shape of the convective plume may differ significantly from the vertical, which is characteristic, for example, water, Prandtl number is $Pr \approx 7$. In this paper we used silicone oil with a density $\rho = 963 \ kg/m^3$ and Prandtl number $Pr \approx 1808$.

Convective plume generated by a local heat source, located on the top of the rubber cylinder, which is located in the center of the bottom of the rectangular cell. To simulate the hot-spot green laser with power $0.2 \ w$ has been used. The angle of incidence ranged from $56°$ to $61°$. A thermocouple placed in a thermal spot showed that the fluid temperature exceeds the average oil temperature in the cavity, which is close to $30°C$. The laser beam enters the cavity through the left vertical edge and slightly warms it. The cell has an inside size 90 x 90 x 80mm. The front and rear walls, through which the observation, the square and made of optical glass, Plexi glas sides and bottom of the plastic. All the walls have a thickness of 10 mm. The cell is filled to the level $83 \ mm$. Under these conditions, the cell is observed in the cellular steady-state flow, whose velocity near the free surface was about $0.05 \ mm/s$.

Visualization was carried out by the schlieren method using device IAB-451 and fixed by a shooting with time interval of $20 \ s$. After switching on the laser plume emerging thin cylinder (see Fig.1), which grows inclined in the direction of flow, breaking the middle cavity height, is set adrift, and then describes a spiral. The height of the oil over the heat source was 7.2 cm, the first turn occurred at a height of 4 cm from the heat source and the upper point of the coil was located at a height of 6cm.

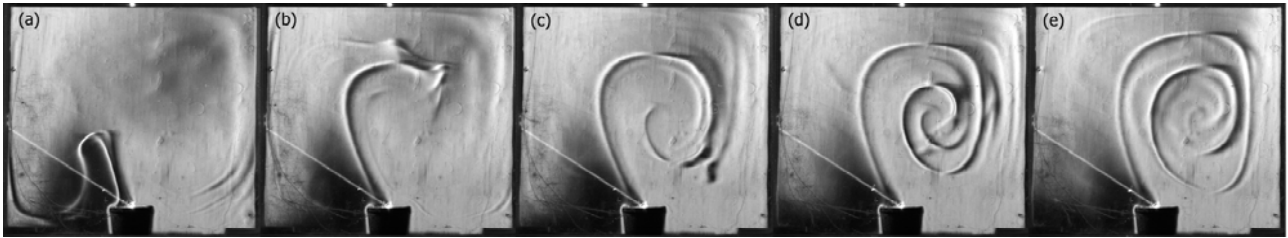

**Fig. 1** The shape of the plume at five time points $t$ after switching laser on: $t = 980s$ (a); $t = 3084s$ (b); $t = 5688s$ (c); $t = 9852s$ (d); $t = 13505s$ (e); The growth rate of the plume was about $0,04$ $mm/s$, Roll type flow has velocity on upper free surface $0,05$ $mm/s$,

It is seen that the upper part of the plume is almost horizontal section. With an increase in turns, this horizontal section is raised and approaches the free surface.

**Conclusion.** In the present experiment showed that the presence of cellular convective motion (simulating the large-scale shear flow exists beneath the plates) the plume from a point source of heat (core hot point) can acquire a spiral shape with horizontal sections needed to launch the mechanism of formation of chains of volcanic islands [3].

**Acknowledgments.** We thank J.A. Whitehead, A.A. Alexeenko and V.I. Stepanov for helpful discussion.